\documentclass[aip,amsmath,smssymb,reprint,graphicx]{revtex4-1}
\usepackage{graphicx}

\draft 

\begin{document}
\newcommand{\rhoEH}{$\rho_{eh}$}
\newcommand{\rhoHH}{$\rho_{hh}$}
\newcommand{\rhoEE}{$\rho_{ee}$}


\title{Switching between attractive and repulsive Coulomb-interaction-mediated drag in
an ambipolar GaAs/AlGaAs bilayer device} 



\author{B.~Zheng}
\author{A.F.~Croxall}
\author{J.~Waldie}
\email[E-mail address: ]{jw353@cam.ac.uk}
\affiliation{Cavendish Laboratory, University of Cambridge, J.J.~Thomson Avenue, Cambridge CB3 0HE, United Kingdom}
\author{K.~Das Gupta}
\affiliation{Cavendish Laboratory, University of Cambridge, J.J.~Thomson Avenue, Cambridge CB3 0HE, United Kingdom}
\affiliation{Department of Physics, Indian Institute of Technology Bombay, Mumbai 400076, India}
\author{F.~Sfigakis}
\affiliation{Cavendish Laboratory, University of Cambridge, J.J.~Thomson Avenue, Cambridge CB3 0HE, United Kingdom}
\author{I.~Farrer}
\altaffiliation[Present address: ]{Department of Electronic and Electrical Engineering, University of Sheffield, Sheffield, S1 3JD, United Kingdom}
\affiliation{Cavendish Laboratory, University of Cambridge, J.J.~Thomson Avenue, Cambridge CB3 0HE, United Kingdom}

\author{H.E.~Beere}
\author{D.A.~Ritchie}
\affiliation{Cavendish Laboratory, University of Cambridge, J.J.~Thomson Avenue, Cambridge CB3 0HE, United Kingdom}


\date{\today}

\begin{abstract}
We present measurements of Coulomb drag in an ambipolar GaAs/AlGaAs double quantum well structure that can be configured as both an electron-hole bilayer and a hole-hole bilayer, with an insulating barrier of only 10 nm between the two quantum wells. The Coulomb drag resistivity is a direct measure of the strength of the interlayer particle-particle interactions. We explore the strongly interacting regime of low carrier densities (2D interaction parameter $r_s$ up to 14). Our ambipolar device design allows comparison between the effects of the attractive electron-hole and repulsive hole-hole interactions, and also shows the effects of the different effective masses of electrons and holes in GaAs.


\end{abstract}


\maketitle 

Bilayer systems consisting of closely spaced two-dimensional (2D) electron or hole gases have attracted intense interest because they are expected to support novel phases stabilised by the interlayer Coulomb interaction, such as an excitonic superfluid, coupled Wigner crystals and charge density waves.\cite{Eisenstein2004, Butov2007,Losovik1976,Swierkowski1991,dePalo2002,Min2008} In particular, the electron-hole bilayer is predicted to form a superfluid coherent state of excitons (electron-hole pairs) at low enough temperature and interlayer separation. Signs of such an excitonic condensate have been observed in optically-generated electron-hole bilayers,\cite{Butov2007,High2012} and in electron-electron or hole-hole bilayers in magnetic fields, where each layer contains a half-filled Landau level of electrons or holes.\cite{Eisenstein2004} However, it has proved very challenging to fabricate stable electron-hole bilayers in zero magnetic field with sufficiently small interlayer separation for the formation of excitonic states while maintaining a sufficiently low interlayer leakage current to avoid electron-hole recombination.

The strength of the interlayer interaction in a 2D bilayer can be probed using Coulomb drag experiments, where a current driven through one layer gives rise to an open-circuit voltage in the other layer, because interlayer interactions transfer momentum from the current-carrying layer to the open-circuit layer.\cite{Gramila1991} As well as the direct interlayer Coulomb interaction, such studies have shown evidence of phonon- and plasmon-mediated\cite{Gramila1993, Hill1997} interactions, and of the effects of particle-particle correlations at low densities.\cite{Kellogg2002,Pillarisetty2002,Yurtsever2003,Hwang2003} Coulomb drag measurements of electron-hole bilayers\cite{Sivan1992} have shown some evidence suggesting a non-Fermi-liquid phase at low temperatures, but these results are not fully understood yet.\cite{Croxall2008,Seamons2009,Gorbachev2012}

In this letter, we report measurements of Coulomb drag in an electron-hole (e-h) bilayer device which can also be operated as a hole-hole (h-h) bilayer. We compare both the magnitude and the density dependence of the electron-hole and hole-hole drag and find that the differences between the two cases can be largely explained based on the different effective masses of the electrons and holes. At the lowest electron and hole densities, the drag shows evidence of significant Coulomb-interaction-driven correlations. Coupled with theoretical modelling, these observations of ambipolar drag should highlight the differences between the effects of attractive e-h and repulsive h-h interactions in bilayer systems.

\begin{figure}[tb]
\includegraphics[width=0.45\textwidth]{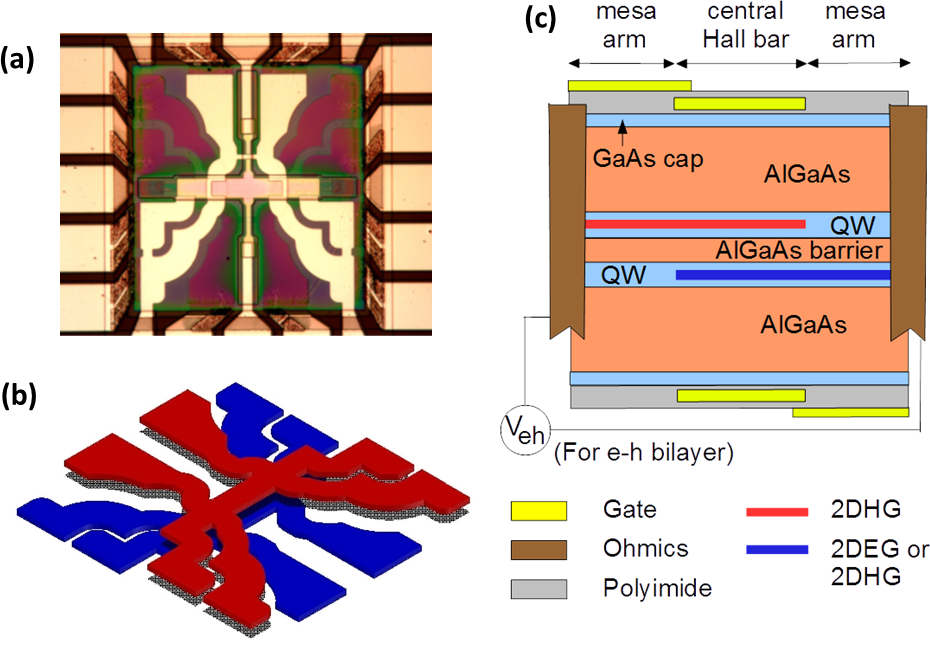}\\
\caption{(Color online) (a) Optical image of the back side of the ambipolar bilayer device; due to the small device thickness ($<1$~$\mu$m), the front-side features can also be seen. (b) Schematic of the geometry of the two overlapping 2D gases. (c) Cross-section view of the device, showing the GaAs/AlGaAs layer structure.}%
\label{fig:device}
\end{figure}
Our device, illustrated in Fig.~\ref{fig:device}, is based on an undoped GaAs/AlGaAs double quantum well (DQW) structure grown on the (100) surface of a semi-insulating (SI) GaAs substrate. Both GaAs quantum wells (QWs) have width 15~nm, and they are separated by a 10-nm-wide Al$_{0.9}$Ga$_{0.1}$As barrier, giving a mean interlayer separation of $d\approx25$~nm. A 2D hole gas (2DHG) can be induced in both quantum wells using negative voltages applied to metallic gates on the front and back of the sample. A positive back-gate voltage can also induce a 2D electron gas (2DEG) in the lower quantum well, when an interlayer bias $V_\text{eh}=1.512$~V is applied between the separate ohmic contacts to the two layers. In order for the back gates to be effective, the SI GaAs substrate is removed, leaving only 300~nm of undoped Al$_{0.33}$Ga$_{0.67}$As and a 10-nm GaAs cap either side of the DQW. We vary the hole density in the front layer, $p_1$, and the electron and hole densities in the back layer, $n_2$ and $p_2$, between $4\times10^{10}$ and $8\times10^{10}$~cm$^{-2}$.
For these densities, the 2DEG interaction parameter $r_{s,e}$ ranges from 2.0 to 2.8, while for the 2DHG $r_{s,h}$ ranges from 10 to 14. $r_{s,e(h)}$ is the ratio of the intralayer e-e (h-h) Coulomb interaction energy to the 2DEG(2DHG) Fermi energy, and is given by $r_{s,e(h)} = m_{e(h)}^* e^2 / 4\pi \epsilon_\text{o}\epsilon_r \hbar^2 \sqrt{\pi n(p)}$, where $m_{e(h)}^*$ is the electron(hole) effective mass and $\epsilon_r = 12.9$ is the GaAs dielectric constant.
The carrier mobilities are in excess of $10^5$~$\text{cm}^{2}\text{V}^{-1}\text{s}^{-1}$ for both electrons and holes for the range of densities studied.

Our device fabrication procedure is similar to that used by Croxall \textit{et al.} to fabricate a symmetrically gated ambipolar single QW structure.\cite{Croxall2013} A Hall bar mesa with twelve arms is defined by chemical etching. P-type ohmic contacts (AuBe alloy, annealed at $500^{\circ}$C for 3 minutes) are placed at the end of each arm, and n-type contacts (AuGeNi alloy, annealed at $470^{\circ}$C for 2 minutes) are also placed at the end of six of the arms. Insulated metal (Ti/Au) gates are patterned over the center of the Hall bar and the six mesa arms that have only p-type contacts. These gates induce the 2DHG in the upper QW. The insulator is a photodefineable polyimide (HD Microsystems HD4104). The SI GaAs substrate is then removed using the epoxy-bond-and-stop-etch procedure.\cite{Weckwerth1996} This process combines mechanical lapping with a three-stage chemical etch, with the thinned sample embedded top-side-down on a thin layer of epoxy and supported by a host substrate. Insulated metal gates are patterned on the back of the sample, overlapping the center of the Hall bar and the six mesa arms that have both n-type and p-type contacts. Depending on the sign of the bias applied to the back-side gates, a 2DEG or 2DHG can be induced in the lower QW. Via holes are etched to allow electrical contact to the ohmic contacts and front-side gates. We note that Seamons \textit{et al.} have previously used a similar procedure to fabricate electron-hole bilayers, but in their samples the polarity of each layer was fixed.\cite{Seamons2007}

For Coulomb drag measurements, the sample is cooled in a sorption-pumped $^3$He cryostat to temperature $T$ between 0.3 and 4.5~K. An a.c. excitation current $I_\text{ex}$ of 10~nA rms at 12~Hz is passed through one layer and the resulting drag voltage $V_\text{D}$ in the other layer is recorded using standard phase-sensitive-detection techniques. We then calculate the drag resistivity $\rho_\text{D} = (V_\text{D}/I_\text{ex})/(l/w)$, where $l/w$ is the length-to-width ratio of the section of the Hall bar between the voltage probes. We performed the standard checks to verify that the measured resistivity is a true Coulomb drag signal.\cite{Gramila1991} The voltage $V_\text{D}$ scales linearly with $I_\text{ex}$ and $(l/w)$. $\rho_\text{D}$ obeys the Onsager reciprocity condition, i.e. it is unaffected by interchanging the current and voltage probes between the layers.\cite{Casimir1945} The interlayer leakage current is less than 100~pA when the sample is biased as an electron-hole bilayer, and far less than this in the hole-hole bilayer case. We have obtained similar results from two samples and on repeated cooldowns of each sample. In the results presented here, we do not see the non-reciprocal upturn of the electron-hole drag that has previously been reported in e-h bilayers.\cite{Croxall2008,Seamons2009}

\begin{figure}[tb]
\includegraphics[width=0.45\textwidth]{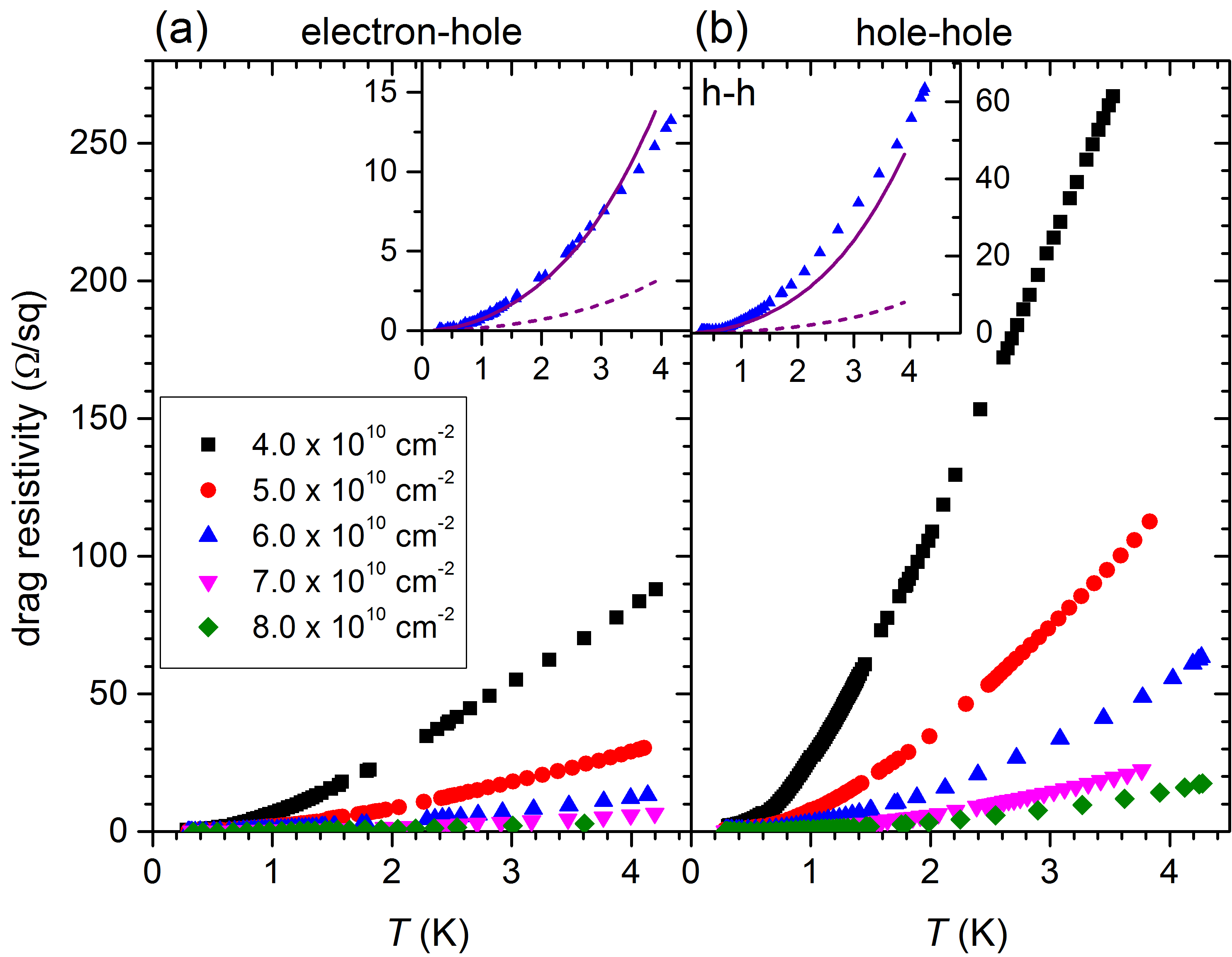}\\
\caption{(Color online) Temperature dependence of the drag resistivities (a) $\rho_\text{eh}$ and (b) $\rho_\text{hh}$, when both layers have equal carrier densities, as given by the legend in (a). Insets compare $\rho_\text{eh}$ and $\rho_\text{hh}$ with theoretical predictions using the RPA (dashed line) and Hubbard (solid line) models.}%
\label{fig:drag_vs_T}
\end{figure}
In Figure \ref{fig:drag_vs_T}, we compare the electron-hole and hole-hole drag resistivities, \rhoEH\ and \rhoHH, for the case of equal carrier densities in both layers. For the range of densities and temperatures studied, we find that \rhoHH\ exceeds \rhoEH\ by factor between 3 and 4. We can also compare our results to those of Kellogg \textit{et al.}, who measured the electron-electron drag (\rhoEE) in a GaAs/AlGaAs device with very similar parameters to ours.\cite{Kellogg2002} The \rhoEH\ data in Fig.~\ref{fig:drag_vs_T} are approximately 4 times larger than the \rhoEE\ results in Ref.~\onlinecite{Kellogg2002}. Thus we have $\text{\rhoHH}>\text{\rhoEH}>\text{\rhoEE}$.

We believe our observation of $\text{\rhoHH}>\text{\rhoEH}$ is linked to the mass asymmetry between electrons and holes in GaAs. While $m_e^* = 0.067m_\text{o}$ ($m_\text{o}$ is the free-electron mass), the effective hole mass is much larger, $m_h^* \approx 5m_e^*$. For Fermi-liquids, in the linear-response regime and to lowest order in the interlayer interaction, the Coulomb drag resistivity is given by\cite{JauhoSmith,Zheng1993}
\begin{equation}
\begin{split}
\rho_\text{D} = &\frac{\hbar^2}{8\pi^2 e^2 n_1n_2k_\text{B}T}
\int^\infty_0 q^3dq \int^\infty_0 d\omega\\ 
&|V_{12}(q,\omega)|^2
\frac{\textrm{Im}[\chi_1(q,\omega)]\textrm{Im}[\chi_2(q,\omega)]}{\sinh^2(\hbar\omega/2k_\text{B}T)},
\end{split}
\label{eqn:BoltzmannDrag}
\end{equation}
where $n_{1(2)}$ is the electron or hole density in layer 1(2), $\chi_{1(2)}(q,\omega)$ is the non-interacting susceptibility of layer 1(2), and $V_{12}(q,\omega)$ is the dynamically-screened interlayer Coulomb interaction. In the limit of low temperature ($T\ll T_{F}$ where $T_{F}$ is the Fermi temperature) and large layer separation ($k_F d \gg 1$, where $k_F$ is the Fermi wavevector), and at high densities ($r_s\lesssim1$) where the Thomas-Fermi theory of screening is valid, Eqn.~(\ref{eqn:BoltzmannDrag}) gives\cite{JauhoSmith}
\begin{equation}
\rho_\text{D} = \frac{\pi^2 \zeta(3)}{16} \frac{h}{e^2}
\left( \frac{\epsilon_\text{o}\epsilon_r k_\text{B}T}{e^2}\right)^2
\frac{1}{(n_1 n_2)^{3/2}d^4},
\label{eqn:ThomasFermi}
\end{equation}
which is independent of the mass of the particles. However, our device is far from this limit, especially in the hole layer. The large hole mass has two consequences. First, the hole Fermi temperature $T_{F,h}$ is much lower than the electron Fermi temperature $T_{F,e}$ at the same density. For our lowest density, $T_{F,e} = 16.4$~K while $T_{F,h} = 3.3$~K. This greatly enhances the effects of thermal excitations in the 2DHG, which weaken the screening of the interlayer Coulomb interaction and therefore enhance the drag. Second, the lower kinetic energy of holes increases the effects of intralayer Coulomb interactions in the hole layers, leading to significant Coulomb-driven correlations between holes in the same layer. These correlations also reduce the ability of the holes to screen the interlayer Coulomb interaction, further increasing the drag.

In the insets to Fig.~\ref{fig:drag_vs_T}, we compare the measured \rhoEH\ and \rhoHH\ with the predictions of Eqn.~(\ref{eqn:BoltzmannDrag}), for $p_1 = p_2(n_2) = 6\times10^{10}$~cm$^{-2}$. Thermal effects are included within the temperature dependence of the susceptibilities $\chi_{1(2)}(q,\omega)$.\cite{Maldague1978} We treat the screened interlayer Coulomb interaction using both the random-phase approximation (RPA, dashed lines) and the zero-temperature Hubbard approximation (HA, solid lines).\cite{Jonson1976}. While the RPA neglects any correlations between particles, the HA accounts for the exchange interaction between particles of the same spin in the same layer, which tends to increase the spacing between such particles. However, neither model incorporates Coulomb-driven intralayer or interlayer correlations. The results in Fig.~\ref{fig:drag_vs_T} show that, for $p_1 = p_2(n_2) = 6\times10^{10}$~cm$^{-2}$, the HA gives a reasonable approximation to the measured drag, while the RPA significantly underestimates it. This is in agreement with previous results.\cite{Pillarisetty2002, Hwang2003, Hwang2008}. Therefore the measured drag is consistent with the electron and hole layers remaining in the Fermi liquid regime. However, it is evident in Fig.~\ref{fig:drag_vs_T} that the HA does not accurately describe the temperature dependence of \rhoEH\ and \rhoHH, and we will now show that it also fails to describe the density dependence.

\begin{figure}[tb]
\includegraphics[width=0.45\textwidth]{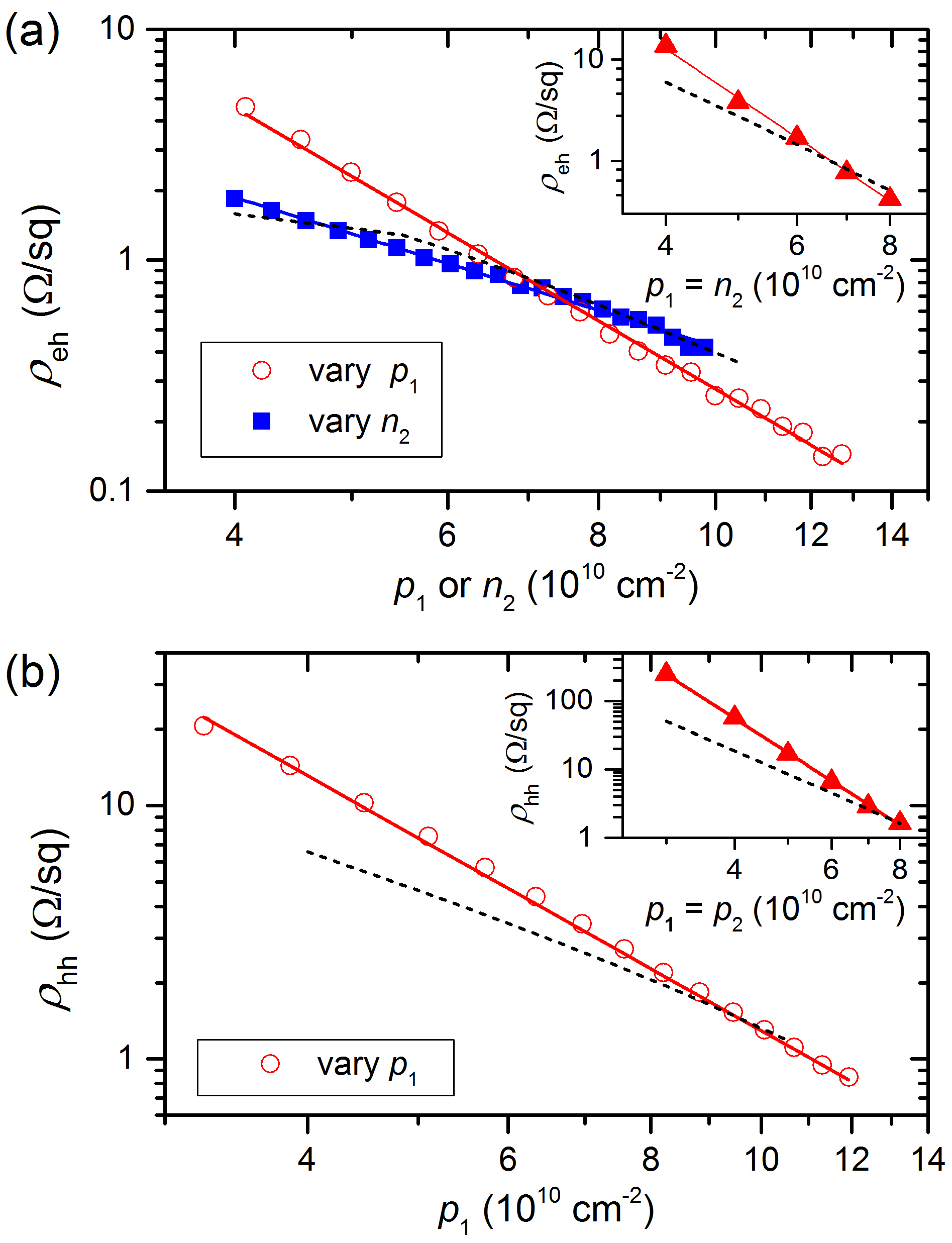}\\
\caption{(Color online) Density dependence of (a) \rhoEH\ and (b) \rhoHH, at $T=1.4$~K. Main plots: the density of one layer is fixed at $7\times10^{10}$~cm$^{-2}$ while the density of the other layer is varied. Insets: both layer densities are kept equal. Solid lines show power-law fits to the experimental data. Dashed lines show the prediction of Eqn.~(\ref{eqn:BoltzmannDrag}) using the HA.}%
\label{fig:drag_vs_np}
\end{figure}

In Fig.~\ref{fig:drag_vs_np}, we explore the density dependence of \rhoEH\ and \rhoHH, by keeping the density of one layer fixed at $7\times10^{10}$~cm$^{-2}$ and varying the density in the other layer. The data in Fig.~\ref{fig:drag_vs_np} are taken at $T=1.4$~K, but we find similar results in the range 0.3 - 2.7~K. In the electron-hole case [Fig.~\ref{fig:drag_vs_np}(a)], we find $\text{\rhoEH}\propto p_1^{\alpha}$ with $\alpha = -3.0$, and $\text{\rhoEH}\propto n_2^{\beta}$ with $\beta_2 = -1.6$, i.e., the drag is much more sensitive to the hole density than the electron density. The inset to Fig.~\ref{fig:drag_vs_np}(a) shows the dependence of \rhoEH\ on $p_1$ (or $n_2$) for matched densities ($p_1=n_2$), and we find $\text{\rhoEH}\propto(p_1=n_2)^{\gamma}$ with $\gamma = -5.0$. The small difference between $\gamma$ and $\alpha+\beta$ is linked to a slight density dependence of the exponents. For \rhoHH\ [Fig.~\ref{fig:drag_vs_np}(b)], we find the same dependence on the densities of both layers, $\text{\rhoHH} \propto (p_1 p_2)^\delta$ with $\delta = -2.65$, in agreement with previous results.\cite{Pillarisetty2002}

The dashed lines in Fig.~\ref{fig:drag_vs_np} show the predicted density dependence based on Eqn.~(\ref{eqn:BoltzmannDrag}), using the zero-temperature HA. This model predicts a very similar dependence of \rhoEH\ and \rhoHH\ on both the electron and hole densities, $\text{\rhoEH}\propto p_1^{-1.7}n_2^{-1.6}$ and $\text{\rhoEH}\propto (p_1 p_2)^{-1.8}$. Therefore, the zero-temperature HA successfully explains the dependence of \rhoEH\ on the electron density $n_2$. This is as expected, because the 2DEG in our system has $T_{F,e}\gg T$ and $r_s \sim 1$. In contrast, both \rhoEH\ and \rhoHH\ depend much more strongly on the 2DHG density than predicted by the zero-temperature HA, because the screening of the interlayer interaction is significantly weakened by thermal excitations and Coulomb-driven intralayer correlations in the hole layers and these effects become more pronounced as $p_1$ and $p_2$ are lowered.

We note that Fig.~\ref{fig:drag_vs_np} shows no evidence of enhancement of either \rhoEH\ or \rhoHH\ for the condition of matched layer densities. Such an enhancement can occur if there is a significant contribution to the drag from the bilayer plasmon modes\cite{Hill1997} or phonon-mediated interactions.\cite{Gramila1993} Therefore we do not believe either mechanism to be significant in our system.

It is clear from our results that Coulomb-driven intralayer correlations in the 2DHG significantly affect the e-h and h-h drag. What is less clear is whether the drag shows evidence of interlayer correlations, which may be a precursor to a transition to a non-Fermi-liquid phase. In the e-h bilayer, the attractive interlayer interaction should lead to increased probability of finding an electron close to a hole and therefore reduced screening of the interlayer interaction and stronger drag. In the h-h bilayer, the repulsive interlayer interaction would have the opposite effect. There is a need for more detailed modelling of our ambipolar system, including the effects of Coulomb-driven intralayer and interlayer correlations, using schemes such as the effective-interaction theories of Refs.~\onlinecite{Yurtsever2003} and \onlinecite{Swierkowski1995}. Comparison of these models with the experimental results should reveal whether the bilayer system is close to the onset of a new phase. 

In conclusion, we have demonstrated measurement of Coulomb drag in an ambipolar device that can be operated as both an electron-hole and a hole-hole bilayer. The hole-hole drag is much stronger than the electron-hole drag, and the drag is more sensitive to the density of the holes than the electrons. We attribute these observations to the large effective hole mass, which makes the effects of thermal excitations and intralayer particle-particle correlations much stronger in the hole layers than the electron layer. We believe a more detailed comparison of ambipolar drag with theoretical models should reveal whether the bilayer system is close to a transition to one of the many predicted non-Fermi-liquid 2D bilayer states.
\\
\par
This work was supported financially by the UK Engineering and Physical Sciences Research Council. AFC acknowledges financial support from Trinity College, Cambridge, and IF from Toshiba Research Europe.

\end{document}